\documentclass{article}
\usepackage{spconf,amsmath,graphicx,adjustbox,hyperref}
\usepackage{color,xcolor}
\usepackage[hang,flushmargin]{footmisc} 
\usepackage{siunitx}
\usepackage{booktabs}
\usepackage{caption}
\usepackage{fancyhdr}
\usepackage{multirow}

\title{Leveraging In-the-Wild Data for Effective Self-Supervised Pretraining in Speaker Recognition}
\name{
\textit{Shuai Wang}$^{1,*}$,
\textit{Qibing Bai}$^{1,2,*}$, 
\textit{Qi Liu}$^3$, 
\textit{Jianwei Yu}$^3$,  
\textit{Zhengyang Chen}$^4$,  \\ 
\textit{Bing Han}$^4$, 
\textit{Yanmin Qian}$^4$,   
\textit{Haizhou Li}$^{1,2}$\thanks{* Co-first author}
}

\address{
\textsuperscript{1}Shenzhen Research Institute of Big Data, Shenzhen, China\\
\textsuperscript{2}The Chinese University of Hong Kong, Shenzhen (CUHK-Shenzhen), China\\
\textsuperscript{3}Tencent, Shenzhen, China \\
\textsuperscript{4}Shanghai Jiao Tong University, Shanghai, China 
}
\fancypagestyle{firstpage}{
    \fancyhf{}
    \fancyfoot[L]{\footnotesize\textcopyright 2024 IEEE. Personal use of this material is permitted. Permission from IEEE must be obtained for all other uses, in any current or future media, including reprinting/republishing this material for advertising or promotional purposes, creating new collective works, for resale or redistribution to servers or lists, or reuse of any copyrighted component of this work in other works.}
 }
\begin{document}
\ninept
\maketitle
\begin{abstract}
Current speaker recognition systems primarily rely on supervised approaches, constrained by the scale of labeled datasets. To boost the system performance, researchers leverage large pretrained models such as WavLM to transfer learned high-level features to the downstream speaker recognition task. However, this approach introduces extra parameters as the pretrained model remains in the inference stage. Another group of researchers directly apply self-supervised methods such as DINO to speaker embedding learning, yet they have not explored its potential on large-scale in-the-wild datasets. In this paper, we present the effectiveness of DINO training on the large-scale WenetSpeech dataset and its transferability in enhancing the supervised system performance on the CNCeleb dataset. Additionally, we introduce a confidence-based data filtering algorithm to remove unreliable data from the pretraining dataset, leading to better performance with less training data. The associated pretrained models, confidence files, pretraining and finetuning scripts will be made available in the Wespeaker toolkit.
\end{abstract}
\begin{keywords}
self-supervised learning, DINO, in-the-wild, speaker recognition
\end{keywords}
\section{Introduction}
\label{sec:intro}

\begin{figure*}[!htb]
    \centering
    \includegraphics[width=0.95\textwidth]{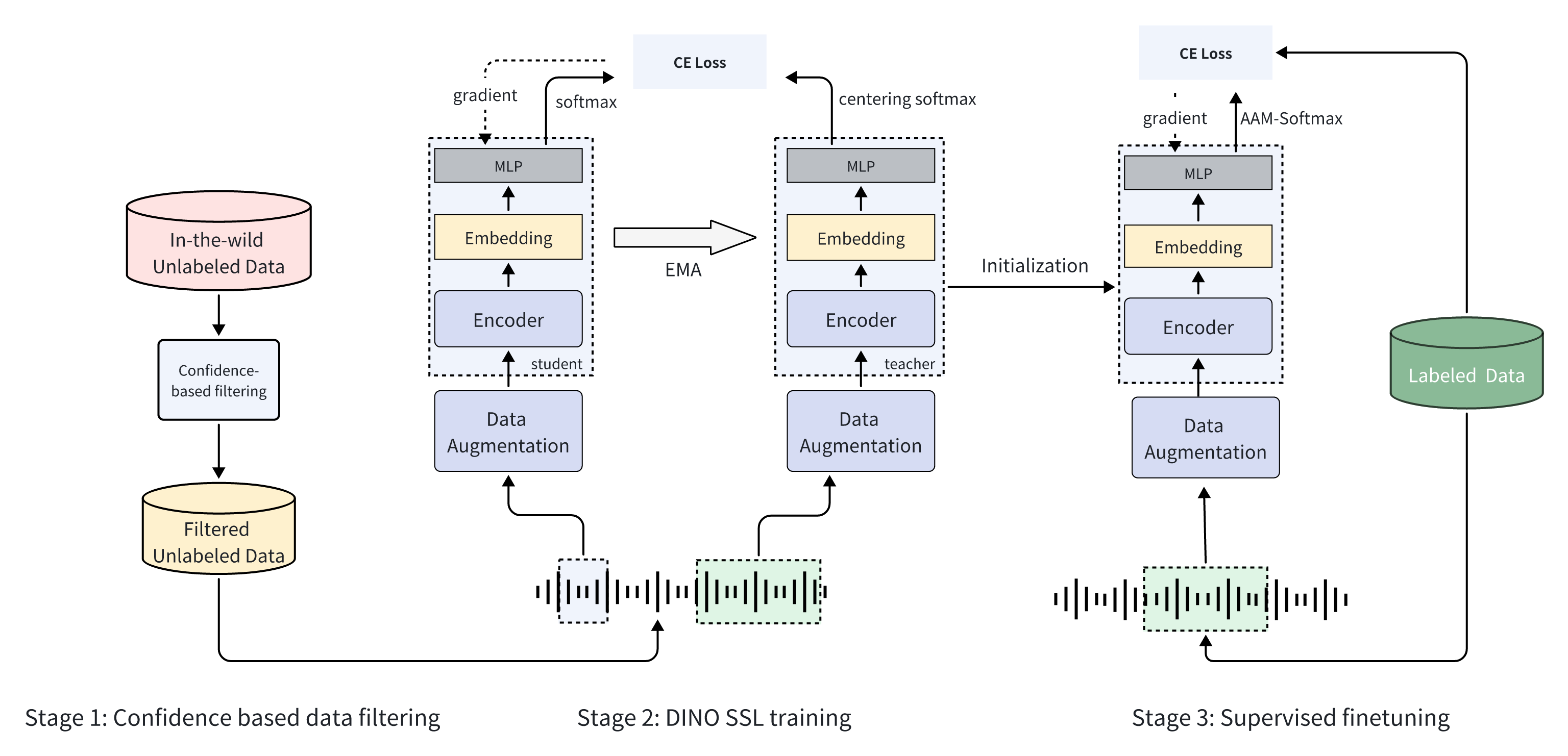}
    \caption{The cascade speaker embedding learning pipeline}
    \label{fig:pipeline}
\end{figure*}

Deep speaker embedding learning plays a central role in applications related to speaker identity modeling, especially in the field of speaker recognition. Current state-of-the-art systems predominantly adhere to the supervised training paradigm, where speaker labels are employed as the optimization target during the training process. Due to the need for annotated data, large-scale datasets like VoxCeleb~\cite{nagrani2017voxceleb,chung2018voxceleb2} and CNCeleb\cite{fan2020cn,li2022cn} with speaker labels have attracted significant attention among researchers. To further enhance the performance of related systems, some researchers have turned to universal speech models pretrained on large-scale in-the-wild data, such as WavLM~\cite{chen2022wavlm}, Wav2Vec~\cite{schneider2019wav2vec} and Hubert~\cite{hsu2021hubert}. They extract higher-level features from these models~\cite{wang2021fine, chen2022large} or utilize them as initialization models for finetuning~\cite{peng2023parameter}. Additionally, some researchers found that using models trained in Automatic Speech Recognition (ASR) as initialization can also improve the performance of speaker recognition systems~\cite{liao2023towards,cai2023pretraining}.

In contrast, researchers have been investigating the integration of various self-supervised training methods for direct training of speaker representation models. Notable methods in this domain include SimCLR~\cite{chen2020simple}, MoCo~\cite{he2020momentum}, and DINO~\cite{caron2021emerging}. Among these approaches, DINO has exhibited particularly impressive performance~\cite{heo2022self,chen2023comprehensive,zhang2022c3}.
However, despite their ability to leverage unlabeled data, current research appears to be primarily focused on pretraining speaker representation models on well-labeled datasets such as VoxCeleb.
There is limited exploration of these methods on unlabeled in-the-wild data, which can be attributed primarily to two factors.
Firstly, our validation indicates that self-supervised models trained on large-scale data do not exhibit strong performance when directly applied to speaker recognition datasets. 
Secondly, these self-supervised learning methods inherently impose certain data requirements that might not be met by in-the-wild datasets. For example, it is crucial for a training segment to contain only one speaker, which necessitates meticulous data cleaning, particularly when dealing with in-the-wild data.

In this paper, we propose a strategy that leverages self-supervised training on large-scale in-the-wild data to initialize supervised speaker models. Our approach offers several key advantages compared with the methods mentioned above: 1) Unlike methods employing large models such as WavLM, our approach introduces no additional parameters or computational overhead during the inference stage, making it more efficient. 2) In contrast to strategies that rely on speech recognition models for initialization, our approach does not require any labels, including training transcripts. 
Our contributions can be summarized as follows:
\begin{itemize}
    \item We introduce a novel learning method that leverages large-scale in-the-wild unlabeled data to significantly enhance the performance of speaker recognition systems. Our approach achieves an overall 12.4\% reduction in Equal Error Rate (EER) on the CNCeleb dataset, exhibiting the immense potential of in-the-wild unlabeled data.
    \item Building upon the consistency assumption of DINO, which assumes the presence of only one speaker in each training segment, we introduce a data filtering technique utilizing confidence scores generated by a speaker diarization system. 
    Through this simple data-cleaning process, we establish that superior pretraining results can be achieved with fewer yet high-quality data, while incorporating more unreliable data does not necessarily improve the performance.

\end{itemize}

\section{Cascade Speaker Embedding Learning}

In this section, we describe our cascade speaker embedding learning pipeline, adept at harnessing large-scale, real-world data to improve the efficacy of supervised training. The complete pipeline is illustrated in Figure~\ref{fig:pipeline} and encompasses three stages: 1) Confidence-based data filtering to acquire high-quality unlabeled data for pretraining. 2) DINO-based pretraining using the filtered unlabeled data. 3) Supervised finetuning using the pretrained DINO model for initialization.

\subsection{Self-supervised Learning on in-the-wild data}
The quality of ``in-the-wild'' datasets is usually diverse. Without manual verification, it becomes challenging to ensure that each segment used in the final training data contains only one speaker. Furthermore, it is crucial to note that increasing the amount of such data does not necessarily guarantee improved results in speaker recognition tasks. The WenetSpeech dataset~\cite{zhang2022wenetspeech} serves as an example highlighting this challenge. Therefore, in this context, we propose a data processing pipeline based on speaker diarization. This pipeline enables effective data filtering, retaining segments of relatively high quality for subsequent pretraining objectives.

\subsubsection{Confidence-based data filtering}

\begin{figure}[ht]
    \centering
    \includegraphics[width=0.48\textwidth]{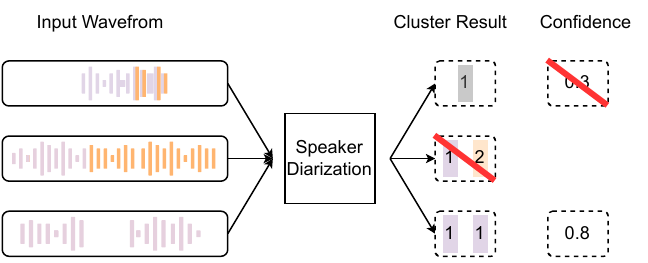}
    \caption{Filtering segments based on diarization results. Colors represent different speakers. Multi-speaker or low-confidence segments will be dropped.} \vspace{-5pt}
    \label{fig:filter}
\end{figure}

The pipeline of the proposed data filtering method is illustrated in Figure~\ref{fig:filter}. Each audio segment in the dataset is processed using a standard clustering-based diarization pipeline, resulting in two possible outcomes: 1) only one speaker is detected, or 2) multiple speakers are present.

For segments falling into the second category, where multiple speakers are present, we simply discard them. However, for segments in the first category, potentially featuring a single speaker, we employ a standard sliding window approach to compute a series of short-chunk embeddings. These embeddings are subsequently compared with corresponding cluster centers to measure their similarity, which is regarded as a form of confidence score.
If the average score is relatively low, it suggests the possibility of either noisy data or unsuccessful separation of multiple speakers. In either scenario, we consider the segment as low-quality and suitable for filtering out.

\subsubsection{DINO Training}

%
As shown in Figure~\ref{fig:pipeline}, the DINO algorithm is implemented using a self-distillation paradigm. The system consists of two sub-networks: the student network and the teacher network, both sharing an identical neural network architecture. During training, following the methodology described in \cite{chen2023comprehensive}, we sample $M$ short segments (local views) and $N$ long segments (global views) from the same utterance. All global and local views are then inputted into the student network, while only global views are inputted into the teacher network. The outputs from the teacher network serve as pseudo labels to guide the optimization of the student network. The optimization loss for the student network can be formulated as follows:
\begin{equation}
\label{eq:dino_loss}
\begin{split}
\mathcal{L}_{DINO} = \frac{1}{N(N+M-1)}
\sum_{i=1}^{N} \sum_{\substack{j=1 \\ j \neq i}}^{N+M}
H(F_t (x^i), F_s (x^j))
\end{split}
\end{equation}
where $x^i$ denotes a global view, $x^j$ denotes a local view, $F_t$ denotes the teacher network, $F_s$ denotes the student network, and $H(\cdot, \cdot)$ represents the cross-entropy loss.

In contrast to the student network, which is optimized using the loss $\mathcal{L}_{DINO}$, the teacher network undergoes updates through the exponential moving average (EMA) applied to the student network. Regarding the architecture of both networks, an initial speaker encoder maps the input to the speaker embedding. This low-dimensional embedding is then processed by several MLP layers to obtain a high-dimensional vector. Finally, a softmax function is applied to this vector, transforming it into a probability distribution. Notably, the teacher network incorporates an additional centering operation before the softmax function.

\subsection{Supervised finetuning}
After stage 2, where we pretrain the model using DINO on unlabeled data, we use this pretrained model as the initial point for our supervised learning stage. It is worth noting that unlike previous studies~\cite{peng2023parameter,peng2023improving}, where additional models or parameters were introduced for adaptation, our approach maintains the exact same architecture while discarding specific prediction layers utilized in stage 2.

\section{Experiments}

\subsection{Dataset}

\textbf{WenetSpeech}: WenetSpeech is an open-source ASR corpus that contains over 10,000 hours of Mandarin data from various sources, including YouTube and Podcasts. Since it is collected from real-world data, WenetSpeech serves as a large-scale in-the-wild dataset and is used for the DINO-based self-supervised pretraining.

\vspace{2mm}\noindent\textbf{CNCeleb}: For the supervised finetuning dataset, we merge the development sets of CNCeleb1 and CNCeleb2 to create the final training set (\textbf{CNCeleb-Train}). The standard CNCeleb test set is used for evaluation (\textbf{CNCeleb-Eval}). Following the Wespeaker recipe\footnote{\url{https://github.com/wenet-e2e/wespeaker/tree/master/examples/cnceleb/v2}}, we concatenate short utterances from the same speaker to ensure that the samples in the training set are longer than 6 seconds. During scoring, we average the embeddings of multiple enrollment utterances to get a single enrollment embedding for each speaker.

\subsection{Experimental settings}
\noindent\textbf{Backbone}: 
 ECAPA-TDNN~\cite{desplanques2020ecapa} is selected as the backbone model for both the DINO pretraining and supervised finetuning. There are 1024 channels in the frame-level convolutional layers, and the final output dimension of DINO is set to 65536.

\vspace{2mm}\noindent\textbf{Training Details}:
The DINO model is trained using the Wespeaker toolkit~\cite{wang2023wespeaker}\footnote{The WenetSpeech training recipe and pretrained models will be released in Wespeaker toolkit.}.
For each training sample, we extract two global views and four local views, and then apply random augmentations to all the sampled segments. Both the student and the teacher receive the global views, each spanning 0.3 seconds in duration. The local views, each lasting 0.2 seconds, are only fed to the student. After obtaining the pretrained model, we finetune it for additional 50 epochs on the CNCeleb dataset, initialized from either the teacher model or the student model.

\vspace{2mm}\noindent\textbf{Evaulation Metrics}:
The cosine back-end serves as the scoring method and the performance of all systems is assessed based on two metrics: Equal Error Rate (EER) and minimum Detection Cost Function (minDCF) with $P_{target}=0.01$.

\vspace{2mm}\noindent\textbf{Data Filtering Configuration}:
First, we obtain segments based on the Voice Activity Detection (VAD) information provided in WenetSpeech. Once we have this information, we follow the speaker diarization recipe\footnote{\url{https://github.com/wenet-e2e/wespeaker/tree/master/examples/voxconverse/v1}} implemented in the Wespeaker toolkit, while the pretrained \textit{ResNet293} model~\footnote{\url{https://github.com/wenet-e2e/wespeaker/blob/master/docs/pretrained.md}} is used as the speaker embedding extractor. For each segment, we perform standard clustering-based diarization. This involves using a sliding window to extract embeddings, followed by applying spectral clustering. After obtaining the clustering results, we eliminate all segments that contain multiple speakers. For the remaining segments, we calculate the cosine similarity between each embedding and its corresponding cluster center, which serves as the confidence score. We retain only those segments with an average score greater than $0.4$. It's worth noting that before these steps, we apply denoising to all the speech data using a denoising tool based on BSRNN~\cite{luo2023music}, considering that WenetSpeech is quite noisy.
After filtering, we discard approximately half of the WenetSpeech dataset, resulting in the differences shown in Table~\ref{tab:datainfo}.
\begin{table}[!htb]
  \centering
  \caption{Statistics of WenetSpeech before and after filtering}
  \label{tab:datainfo}
      \begin{adjustbox}{width=0.45\textwidth,center}
  \begin{tabular}{lcccc}
    \toprule
    Dataset & Number of Segments & Total Duration \\
    \midrule
    WenetSpeech & 17,848,005 & 12483.35h \\
    Filtered WenetSpeech & 9,661,524 & 6816.43h \\
    \bottomrule
  \end{tabular}
  \end{adjustbox}
\end{table}

\subsection{Results and Analysis}

\subsubsection{Baseline system}
To establish a strong baseline, we incorporated large-margin finetuning~\cite{thienpondt2021idlab} and adaptive symmetric normalization (AS-norm)~\cite{matvejka2017analysis} into our pipeline. This resulted in a noticeable performance improvement, as indicated in Table~\ref{tab:result_baseline}. The same processes will be applied to all the systems in the subsequent context.

\begin{table}[ht]
  \centering
  \caption{Performance of the baseline system trained on CNCeleb-Train and evaluated on CNCeleb-Eval}
  \label{tab:result_baseline}
  \begin{adjustbox}{width=0.4\textwidth,center}
  \begin{tabular}{lcc}
    \toprule
     System & EER (\%) & MinDCF \\
    \midrule
    Baseline & 7.879 & 0.420 \\
    + AS-norm & 7.412 & 0.379  \\
    ++ Large-margin finetuning & \textbf{7.395} & \textbf{0.375}  \\
    \bottomrule
  \end{tabular}
  \end{adjustbox}
  \vspace{-2mm}
\end{table}

\begin{table*}[!htb]
  \centering
  \caption{Comparison of performance on CNCeleb-Eval with other pretrain-finetune methods. CNCeleb-Train contains both CNCeleb1 and CNCeleb2 as defined in Section 3.1.}
  \label{tab:comparison}
  \begin{adjustbox}{width=0.96\textwidth,center}
  \begin{tabular}{c|ccc|cc|cc}
    \toprule
    \multirow{2}{*}{System} & \multicolumn{3}{c|}{Pretraining Configurations}&\multicolumn{2}{c|}{Finetuning Configurations}&\multirow{2}{*}{EER(\%)}  & \multirow{2}{*}{MinDCF} \\ \cmidrule(r){2-4}\cmidrule(r){5-6}
    & Data & Model & Role  &  Data &  Model & \\
    \midrule
    \cite{heo2022self} & VoxCeleb2 & ECAPA-TDNN & Init& CNCeleb1 & ECAPA-TDNN & 10.65 & - \\

    \cite{kang2022augmentation} & VoxCeleb2 & ECAPA-TDNN & Init & CNCeleb1 & ECAPA-TDNN & 8.710 & 0.422 \\
    \cite{han2023improving} & VoxCeleb2 & ECAPA-TDNN & Init & CNCeleb1 & ECAPA-TDNN & 10.03 & 0.539 \\
    \cite{peng2023improving} & CNCeleb1 & HuBERT (94.6M) & Frontend &  CNCeleb1 & HuBERT + ECAPA-TDNN  & 10.86 & - \\
    \cite{peng2023improving} & CNCeleb-Train & HuBERT (94.6M) & Frontend &  CNCeleb-Train & HuBERT + ECAPA-TDNN  & 8.890 & - \\
    \cite{peng2023improving} & CNCeleb-Train & Conformer (172.2M) & Frontend &  CNCeleb-Train & Conformer + MHFA  & 7.730 & 0.406 \\
    \cite{peng2023parameter} * & Mix 94k hr & WavLM (94.7M) & Frontend & VoxCeleb2 + CNCeleb-Train & WavLM+MAM+MHFA & 6.890 & 0.378 \\
    \cite{liao2023towards}** & WenetSpeech & Conformer (18.8M)  &  Init & CNCeleb-Train & Conformer & 7.420 & 0.443 \\\midrule
   Ours & WenetSpeech & ECAPA-TDNN  & Init  & CNCeleb1 & ECAPA-TDNN & 7.373 & 0.383 \\
    Ours & + filtering  &ECAPA-TDNN & Init & CNCeleb1  & ECAPA-TDNN & 7.339 & 0.377 \\
    Ours & WenetSpeech & ECAPA-TDNN  & Init  & CNCeleb-Train & ECAPA-TDNN & 6.738 & 0.338 \\
    Ours & + filtering  &ECAPA-TDNN & Init & CNCeleb-Train  & ECAPA-TDNN & \textbf{6.474} & \textbf{0.331} \\
    \bottomrule
  \end{tabular}
  \end{adjustbox}
  \caption*{* The publicly available WavLM Base+ checkpoint is used, which has been pretrained on a significantly larger dataset (94k hours). \\
  ** The pertaining process requires transcriptions because it operates within the ASR framework.
  }
\end{table*}

\subsubsection{Effectiveness of the proposed pipeline}
The performance of the pretrained DINO models is first evaluated considering different training data, as demonstrated in Table~\ref{tab:result_dino}. In the literature, it has not been explicitly mentioned whether to employ the teacher or student model for evaluation and finetuning~\cite{han2023improving,heo2022self}. Considering the siamese architecture of DINO, both teacher and student networks are assessed.


\begin{table}[ht]
  \centering
  \caption{Performance comparison of self-supervised DINO models without and with finetuning}
  \label{tab:result_dino}
        \begin{adjustbox}{width=0.48\textwidth,center}
  \begin{tabular}{llcccc}
    \toprule
    Pretraining  & Finetuning & \multicolumn{2}{c}{DINO Teacher} & \multicolumn{2}{c}{DINO Student} \\
    \cmidrule(r){3-4}\cmidrule(r){5-6}
    Data & Data & EER (\%) & MinDCF & EER (\%) & MinDCF \\
    \midrule
    CNCeleb-Train & - & 13.74 & 0.563 & 14.11 & 0.576 \\ 
    WenetSpeech & - & 15.40 & 0.605 & \textcolor{gray}{15.22} & 0.625 \\ 
    + Filtering & - & 15.03 & 0.560 & \textcolor{gray}{15.87} & 0.585 \\\midrule
     CNCeleb-Train & CNCeleb-Train & 7.339 & 0.366 & 7.378 & 0.364 \\
    WenetSpeech & CNCeleb-Train & 6.738 & 0.338 & 6.815 & 0.341  \\
    + filtering & CNCeleb-Train & \textbf{6.474} & \textbf{0.331} & 6.528 & 0.331 \\
    \bottomrule
  \end{tabular}
  \end{adjustbox}
  \vspace{-4mm}
\end{table}

From Table~\ref{tab:result_dino}, we observed that the teacher model exhibits better stability in performance compared to the student model, which aligns with the findings in the paper~\cite{tarvainen2017mean}. This observation might be attributed to the teacher model updating its parameters through the exponential moving average (EMA) of the student model.

Performance comparison between the systems that incorporate models pretrained on WenetSpeech, with and without filtering, exhibits the effectiveness of our proposed data filtering strategy. Notably, better results are achieved with nearly half the training data, as evidenced by improvements in both EER and DCF. This enhancement remains consistent regardless of whether we employ the teacher or student model for initialization. In comparison with the baseline, our training pipeline decreases the EER from 7.395\% to 6.474\%, achieving a relative reduction of 12.45\%.

\subsubsection{The effect of self-pretraining}
In this section, we evaluate the finetuning performance on CNCeleb when leveraging different self-supervised pretraining models and present the results in Table~\ref{tab:result_dino}.
First, similar to the self-pretraining approach discussed in~\cite{peng2023improving}, we conducted both pretraining and finetuning on the CNCeleb dataset. However, we observed only marginal performance improvement compared to the baseline results presented in Table \ref{tab:result_baseline}.
Additionally, it is noteworthy that the model pretrained on WenetSpeech without data filtering surpasses the self-pretrained model, even though the self-pretrained model is initially the top-performing one prior to finetuning.

\subsubsection{Comparison with other pretraining methods}
To provide a more intuitive demonstration of the effectiveness of our method compared to other pretraining approaches, we summarize a list of results reported in the literature for CNCeleb in Table~\ref{tab:comparison}. This list includes details about the pretraining data and the adaptation/finetuning methods employed. The role the pretrained model plays can be classified into two categories: initialization point and feature extraction front-end. For the former category, the finetuning and inference stages do not introduce any new parameters and maintain the backbone exactly the same. However, for the latter one, adaptation layers or a common speaker embedding back-end must be added to the pretrained model.
Considering that some of the comparison systems are finetuned only using CNCeleb1, we also present our systems under the same setup to ensure a fair comparison.

From the results, we can observe that our top-performing system surpasses other systems that adhere to the pretrain-finetune approach, even though they may incorporate more pretraining data, additional parameters, or require other label types like ASR transcription. Moreover, it is worth noting that our system, which is only finetuned on the small CNCeleb1 dataset, outperforms the supervised baseline trained on the larger CNCeleb-Train dataset (7.339\% versus 7.395\% in terms of EER).

\section{Conclusion}
\label{sec:conclu}
In this paper, we present a pipeline that is able to take advantage of extensive data in the wild to boost the performance of a speaker recognition system. In contrast to prior research that tunes DINO using limited-scale labeled datasets or introduces pretrained models with additional computational cost, we demonstrate that the DINO pretraining on large data can benefit speaker recognition without introducing any extra parameters. Through an efficient data filtering method based on automatic confidence assessment, the proposed system achieves superior results while utilizing only half of the pretraining data.



\bibliographystyle{IEEEbib}
\bibliography{strings,refs}

\end{document}